\documentclass[fleqn,twoside]{article}
\usepackage{espcrc2}

\usepackage{graphicx}
\usepackage[figuresright]{rotating}

\newcommand{\AmS}{{\protect\the\textfont2
  A\kern-.1667em\lower.5ex\hbox{M}\kern-.125emS}}

\title{
  Effects of low-lying fermion modes in the
  $\epsilon$-regime\addressmark\thanks{
    Talk presented by K. Ogawa.} 
}
\author{
  Kenji Ogawa\address[GUAS]{
    Department of Particle and Nuclear Physics, 
    The Gradute University for Advanced Studies,\\ 
    Tsukuba, Ibaraki 305-0801, Japan}
  and
  Shoji Hashimoto\address{
    High Energy Accelerator Research Organization (KEK),
    Tsukuba, Ibaraki 305-0801, Japan}
}

\begin{document}

\begin{abstract}
  We investigate the effects of low-lying fermion modes on
  the QCD partition function in the $\epsilon$-regime. 
  With the overlap Dirac operator we calculate several tens
  of low-lying fermion eigenvalues on the quenched lattice.
  By partially incorporating the fermion determinant through
  the truncated determinant approximation, we calculate the
  partition function and other related quantities for $N_f$
  = 1 and compare them with the theoretical predictions
  obtained by Leutwyler and Smilga.
\end{abstract}

\maketitle

\section{Introduction}
The low energy behavior of QCD, and therefore lattice QCD,
is effectively described by the chiral Lagrangian.
Approaching the massless limit, the pion's Compton wave
length becomes longer than the spatial extent of the
lattice, and the zero momentum modes dominate the partition
function.
This is the so-called $\epsilon$-regime of QCD, where
various properties of the QCD partition function is known
analytically \cite{Leutwyler:1992yt}.
For instance, for one-flavor QCD the partition function
is given by $Z=e^{m\Sigma V\cos\theta}$ for finite $\theta$
at the leading order.
Here, $m$ is the quark mass, $\Sigma$ is the chiral
condensate, and $V$ is the space-time volume.
It can be divided into topological sectors as
\begin{equation}
  Z_\nu(m) = \int\! d\theta\, e^{i\nu\theta} Z 
  = I_{\nu}(m\Sigma V),
\end{equation}
where $I_{\nu}(x)$ denotes the modified Bessel function, and
$\nu$ is the topological charge.

For lattice QCD to study the small quark mass region, it is
therefore the first step to confirm such analytic results.
Eventually one can determine the low energy constants of
chiral perturbation theory by measuring correlation
functions in the $\epsilon$-regime, as discussed for example
in \cite{Damgaard:2001js}.
In this exploratory study we investigate the mass dependence
of the partition function, the sum rules for eigenvalues of
the Dirac operator, as well as the topological
susceptibility.
The effect of low-lying fermion eigenmodes is incorporated
using the reweighting technique, and thus we study the above
relations in the one-flavor $N_f$ = 1 case.

Since the chiral symmetry is essential in deriving the
partition function in the $\epsilon$-regime, we employ the
overlap Dirac operator
$D=\frac{1+s}{a}\left[1+\gamma_5{\mathop{\rm sgn}}(H_W)\right]$.
To compute the sign function ${\rm sgn}(H_W)$,
14 lowest eigenmodes of $H_W$ are treated exactly and 
we used the Chebyshev polynomial of degree 100-200 to
approximate the rest of the eigenmodes.
On a $10^4$ lattice at $\beta$ = 5.85 we calculated 50
lowest eigenvalues of $D$ using the implicit restarted
Arnoldi method (ARPACK). 
The topological charge is obtained from the number of
zero modes for each gauge configuration, and we accumulated
168, 290, and 149 configurations for $|\nu|$ = 0, 1, and 2,
respectively.

\section{Truncated determinant approximation}
In order to study the effect of fermion determinant, we use
an approximation to include only the low-lying fermion
eigenmodes.
Namely, the fermion determinant is replaced by a product
$\prod_{i = 1}^{N_{\max}} (|\lambda_i|^2 + m^2)$,
where $\lambda_i$ is the $i$-th lowest eigenvalue of
the overlap Dirac operator $D$ for a given configuration.
If $N_{\max}$ is large enough to cover physically relevant
eigenmodes, this approximation is expected to reproduce the
right physics.
For our choice of parameters $\lambda_{10}$ lies around
700~MeV, and therefore should already be large enough to be
treated as the ultraviolet mode.
The neglected eigenmodes are the ultraviolet modes, which
should be irrelevant to the low energy physics.
Such idea was implemented previously as an algorithm for the
Wilson fermion \cite{Duncan:1998gq}, in which the
ultraviolet modes are also incorporated using either the
effective gauge action or the multi-boson method.
See also \cite{Borici:2002rq}.

\begin{figure}[tb]
  \centering
  \scalebox{0.5}{\includegraphics{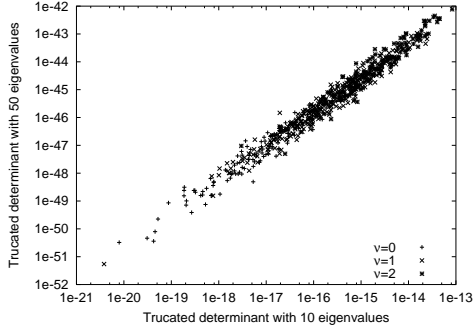}}
  \vspace*{-8mm}
  \caption{
    Correlation between $\prod_{i=1}^{10} |\lambda_i|^2$
    and $\prod_{i=1}^{50} |\lambda_i|^2$ for the topological
    sector 0, 1, and 2.
    For the non-zero topological charge, the zero modes are
    subtracted.
  }
  \label{fig:trdt1}
\end{figure}

\begin{figure}[tb]
  \centering
  \scalebox{0.5}{\includegraphics{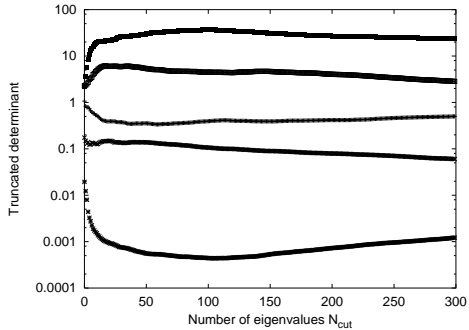}}
  \vspace*{-8mm}
  \caption{
    Truncated determinant as a function of $N_{\max}$,
    normalized with an average over five configurations
    shown in the plot (all in the $\nu$ = 0 sector).
  }
  \label{fig:trdt2}
\end{figure}

In order to demonstrate how the truncation works, in
Fig.~\ref{fig:trdt1} we plot a correlation between the
truncated determinants with $N_{\max}$ = 10 and 50 for 
$m$ = 0. 
Each dot in the plot represents a gauge configuration, 
and the strong correlation among the different truncation 
indicates that the product 
$\prod_{i=11}^{50} |\lambda_i|^2$
is mostly independent of gauge configuration.
In Fig.~\ref{fig:trdt2} we show how the truncated
determinant varies as a function of $N_{\max}$.
We can observe a rapid change below $N_{\max}\sim$ 10, with
which the order of magnitude is determined.
The values still changes slowly below $N_{\max}\sim$ 50, and
stays almost flat above it.

\section{Partition function}

\begin{figure}[tb]
  \centering
  \scalebox{0.5}{\includegraphics{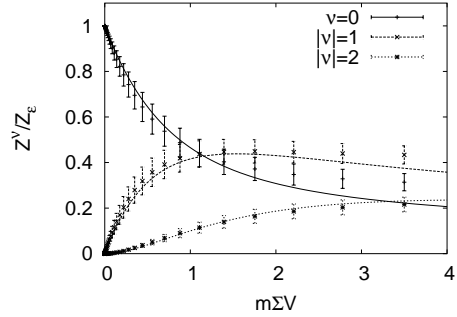}}
  \vspace*{-8mm}
  \caption{
    Partition function of a given topological sector divided
    by the sum of the partition function of all topological
    sectors.
    The curves are the analytic prediction
    \cite{Leutwyler:1992yt}. 
}
\label{fig:detmass_d8}
\end{figure}

In Fig.~\ref{fig:detmass_d8} we show the mass dependence of
the partition function $Z_{\nu}(m)/Z(m)|_{\theta=0}$
for the topological sectors $|\nu|$ = 0, 1, and 2 at 
$N_f$ = 1.
The truncated determinant with $N_{\max}$ = 50 is used.
We can see that the data agree with the analytic expectation
$I_\nu(m\Sigma V)/\exp(m\Sigma V)$ \cite{Leutwyler:1992yt}
shown by the curves.
From this fit we obtain $\Sigma$ = (243 MeV)$^3$.

\begin{figure}[tb]
  \centering
  \scalebox{0.5}{\includegraphics{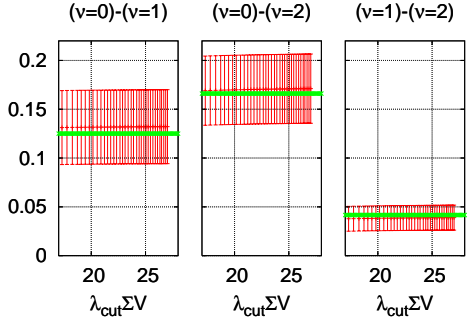}}
  \vspace*{-8mm}
  \caption{
    Difference of 
    $\langle \sum_n 1/(\lambda_n\Sigma V)^2\rangle_\nu$
    among different topological sectors.
    Analytical expectation is shown by horizontal lines. 
  }
  \label{fig:sum_rule}
\end{figure}

By differentiating the partition function with respect to
the quark mass, Leutwyler and Smilga obtained a sum rule for
the eigenvalues \cite{Leutwyler:1992yt}
\begin{equation}
  \left\langle 
    \sum_n \frac{1}{(\lambda_n\Sigma V)^2}
  \right\rangle_\nu 
  =  \frac{1}{4(\nu+1)}.
  \label{eq:sum_rule}
\end{equation}
Left hand side of this equation is quadratically divergent
in the ultraviolet region, since the eigenmode distribution
behaves as $\lambda^3$ in the bulk.
Therefore the sum rule (\ref{eq:sum_rule}) makes sense only
after the ultraviolet divergence is regulated and the
particular volume dependence is extracted.
Here, instead, we consider a difference of
(\ref{eq:sum_rule}) between different topological sectors.
Because the bulk distribution is expected to be independent
of topology, the ultraviolet divergence cancels in such
differences.
Fig.~\ref{fig:sum_rule} shows these differences as a
function of the ultraviolet cutoff.
We find that the results are completely independent of the
cutoff and agree with the analytical predictions.

\section{Topological susceptibility}
\begin{figure}[tb]
  \centering
  \scalebox{0.5}{\includegraphics{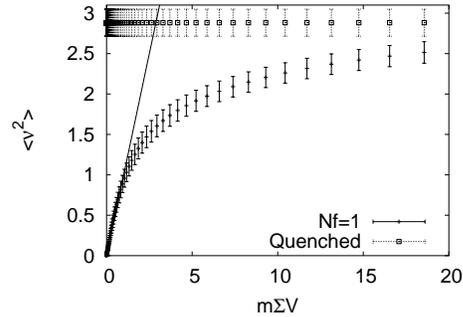}}
  \vspace*{-9mm}
  \caption{
    Mass dependence of the topological charge squared 
    $ \langle \nu^2 \rangle $ for $N_f$ = 0 and 1.
  }
  \label{fig:tpsus_a}
\end{figure}

Topological susceptibility $\chi$ is
obtained from the partition function by differentiating with
respect to $\theta^2$.
On the lattice one can simply calculate the expectation
value of the topological charge squared
$\langle\nu^2\rangle=\chi V$.
In Fig.~\ref{fig:tpsus_a} we plot $\langle\nu^2\rangle$ as a
function of quark mass for both $N_f$ = 0 and 1.
Near the massless limit the topological susceptibility grows
linearly for $N_f$ = 1 as expected from the form of the
partition function $e^{m\Sigma V\cos\theta}$.
For larger quark masses it saturates toward the quenched
value corresponding to $\chi$ = (205 MeV)$^4$.
This behavior was previous observed in \cite{Kovacs:2001jr}.
By fitting the slope near the massless limit we obtain
$\Sigma$ = (238~MeV)$^3$, which agrees well with the
determination from the partition function itself.

\section{Conclusions}
Using the truncated determinant approximation of the overlap
Dirac operator, we confirmed that the Leutwyler-Smilga's
analytic predictions for the QCD partition function is
satisfied for $N_f$ = 1.
Results are stable against the number of eigenmodes
included, if it is greater than 10 
($\lambda_{10} \sim 700{\rm MeV}$).
It suggests that the dynamical fermion algorithms designed
to separate low and high eigenmodes, such as
\cite{Duncan:1998gq}, work effectively.
For larger volume or smaller lattice spacing the number of
eigenvalues to cover the equivalent physical region becomes
much larger (proportional to the lattice volume), and their
calculation could be numerically prohibitive.
Still it would be helpful to isolate the low-lying modes, as
they account for the dominant part of the fermion
determinant.

\end{document}